\documentstyle[12pt]{article}

\let\ssection=\section \def\section{\setcounter{equation}{0}\ssection}

\font\openbig=msym10 scaled\magstep1
\font\openscr=msym7 scaled\magstep1
\font\openscrscr=msym5 scaled\magstep1
\newfam\openfam
\textfont\openfam=\openbig
\scriptfont\openfam=\openscr
\scriptscriptfont\openfam=\openscrscr
\def\open{\fam\openfam}

\font\Scbig=cmss10 scaled\magstep1
\font\Scscr=cmss8 scaled\magstep1
\font\Scscrscr=cmss8
\newfam\Scfam
\textfont\Scfam=\Scbig
\scriptfont\Scfam=\Scscr
\scriptscriptfont\Scfam=\Scscrscr
\makeatletter

\newdimen\normalarrayskip              
\newdimen\minarrayskip                 
\normalarrayskip\baselineskip
\minarrayskip\jot
\newif\ifold             \oldtrue            \def\new{\oldfalse}
\def\arraymode{\ifold\relax\else\displaystyle\fi} 
\def\@arrayskip{\ifold\baselineskip\z@\lineskip\z@
     \else
     \baselineskip\minarrayskip\lineskip2\minarrayskip\fi}
\def\@arrayclassz{\ifcase \@lastchclass \@acolampacol \or
\@ampacol \or \or \or \@addamp \or
   \@acolampacol \or \@firstampfalse \@acol \fi
\edef\@preamble{\@preamble
  \ifcase \@chnum
     \hfil$\relax\arraymode\@sharp$\hfil
     \or $\relax\arraymode\@sharp$\hfil
     \or \hfil$\relax\arraymode\@sharp$\fi}}
\def\@array[#1]#2{\setbox\@arstrutbox=\hbox{\vrule
     height\arraystretch \ht\strutbox
     depth\arraystretch \dp\strutbox
     width\z@}\@mkpream{#2}\edef\@preamble{\halign \noexpand\@halignto
\bgroup \tabskip\z@ \@arstrut \@preamble \tabskip\z@ \cr}%
\let\@startpbox\@@startpbox \let\@endpbox\@@endpbox
  \if #1t\vtop \else \if#1b\vbox \else \vcenter \fi\fi
  \bgroup \let\par\relax
  \let\@sharp##\let\protect\relax
  \@arrayskip\@preamble}
\makeatother
\def\Sc{\fam\Scfam}

\def\lvm{\leavevmode\hbox to\parindent{\hfill}}

\def\req#1{(\ref{#1})}
\def\dd#1{{\partial\over\partial{#1}}}
\def\L{\left}
\def\R{\right}

\def\sumr{\sum_{r\geq1}}
\def\sumji{\sum_{j\neq i}}

\def\BE{\begin{equation}}
\def\EE{\end{equation}}
\def\BA{\begin{array}}
\def\EA{\end{array}}

\def\a{\alpha}
\def\half{{1\over2}}
\def\d{\partial}
\def\dd#1{{\partial\over\partial{#1}}}
\def\ddsc#1{{\partial^2\over\partial{#1}^2}}
\def\w#1{${\rm W}^{({#1})}$}
\def\pr{^\prime}

\def\de{decoupling equation}
\def\h{hierarchy}
\def\hs{hierarchies}
\def\cs{constraints}
\def\K{Kontsevich}
\def\M{Miwa}
\def\V{Virasoro}
\def\mm{matrix model}
\def\cft{conformal field theory}
\def\emt{energy--momentum tensor}

\begin{document}
\hfuzz=1pt

\title{
{\sc Minimal Models from \\ W-Constrained Hierarchies \\ via
the \K--\M\ Transform}}

\author{{\large {\bf B.~Gato-Rivera}}\\
{\small {\sl CERN, CH1211 Geneva 23, Switzerland}}\\
{\small and}\\
{\small {\sl Instituto de F\'\i sica
 Fundamental, Serrano 123, Madrid 28006,
Spain}}\\{}\\ and\\ {}\\
{\large {\bf A.~M.~Semikhatov}}\\ {\small
{\sl Theory Division, P.~N.~Lebedev
Physics Institute}}\\ {\small {\sl Leninsky
prosp. 53, Moscow 117924, Russia}}}

\date{\relax{}}

\maketitle

\begin{abstract}
A direct relation between the
conformal formalism for 2$d$-quantum gravity
and the W-constrained KP \h\ is found, without the need to invoke
intermediate \mm\ technology.  The \K-\M\ transform
of the KP \h\ is used to
establish an identification between W constraints on
the KP tau function and \de s
 corresponding to \V\ null vectors. The \K-\M\ transform maps the
\w{l}-constrained KP \h\ to the $(p\pr,p)$ minimal model, with the tau
function being given by the correlator
of a product of (dressed) $(l,1)$ (or
$(1,l)$) operators, provided the Miwa
parameter $n_i$ and the free parameter
(an abstract $bc$ spin) present in
the \cs\ are expressed through the ratio
$p\pr/p$ and the level $l$.\end{abstract}

\thispagestyle{empty}

\vskip120pt

\noindent
CERN.TH-6469/92

\noindent
IFF-92/4

\noindent
hep-th@xxx/9204085

\noindent
April 1992

\newpage
\section{Introduction}\lvm
Matrix models, which appeared as
a `discretized' approach to 2$d$ quantum gravity
\cite{[BK],[DSh],[GM]}, have resulted at
the continuum level in integrable
\hs\ subjected to \V\ \cite{[FKN1],[DVV]} and possibly higher W
\cite{[DVV],[GaNa]} \cs. However, their
relation with the Liouville theory,
at least with that described
in the formalism of \cite{[Da],[DK]}, seems,
although hardly disputable in principle,
somewhat obscure. In a recent paper
\cite{[S35]} the \K-\M\ transform was
used to establish a relation between
the \V\ \cs\ imposed on the tau
functions of the KP hierarchy, and the \de\
corresponding to the null vector at level 2 in minimal conformal field
theories extended by a scalar
current\footnote{The effect of the \K-\M\
transformation was to replace the time parameters
of the \h\ with the spectral
parameter, or, in different words, to
create a world-sheet out of the time
parameters.}. Moreover, the field content of the David-Distler-Kawai
formalism for 2$d$ quantum gravity was
recovered in this way, with the extra
scalar playing a r\^ole similar to that of the Liouville field in the
formalism of \cite{[Da],[DK]}. The essence of
the method based on the \K-\M\
transform was therefore to relate the
\V-constrained KP \h\ and $2d$ quantum
gravity, bypassing the {\it matrix\/} model technology.

 Now the question arises as to
 whether the higher W \cs\ on the KP \h\
are amenable to the above scheme, and
whether a bridge can be
established between matter interacting with the
continuum DDK quantum gravity and the W-constrained KP \hs.

In this letter we give a positive
answer to this question, based on the
results obtained for the \w3 constraints
in addition to the previous results
for the \w2 (\V) case.

By analogy with the level-2/Virasoro case,
one could expect that under the
\K-\M\ parametrization of the KP times,
the higher ($l\geq3$) W \cs\ ${\Sc
W}^{(l)}_n\tau=0$, $n\geq-l+1$, would lead
to higher-level ($l\geq3$) null
vector \de s\footnote{A relation between \V\
null vectors and W algebras has
been noted in a different approach
in \cite{[BDfIZ]} and, more recently, in
 \cite{[DPW]}.}.
This argument, however, is fraught with an unpleasant
``paradox": the algebra of the
\w{l} constraints contains the lower \w{l\pr},
$l\pr<l$ as well, which apparently should then give rise, by the same
mechanism, to \de s at the lower levels, thereby leading to an
over-determined system. Fortunately, this does
not happen; it turns out that
a single level-$l$ \de\ corresponds to the complete set of \w{l} \cs\
(including the \V\ ones)
${\Sc W}^{(l\pr)}_n\tau=0$, $n\geq-l\pr+1$, $2\leq
l\pr\leq l$, while these do {\it not\/}
lead back to lower-level \de s. More
precisely, the \K-\M\ transform maps
the \w{l}-constrained KP \h\ to the
$(p\pr,p)$ minimal model, with the tau
function being given by the correlator
of a product of (dressed) $(l,1)$
(or $(1,l)$) operators, provided (i) the
Miwa parameter $n_i$ satisfies
$$n_i^2=(l-1)^2{p\pr\over2p},$$ and (ii) the
free parameter (an abstract $bc$ spin)
$J$ present in the \cs\ is determined
from $$2J-1={l-1\over n_i}-{2n_i\over l-1}.$$
Although we have obtained the
complete results only for the \w3
case (in addition to \w2~=~Virasoro), it is
hard to believe that so
simple a result, the W --- decoupling correspondence,
following after rather tedious calculations,
could be a mere coincidence and
would not allow generalizations to \w{\geq4} algebras.

We would like to stress that, rather
than taking the \de\ in its na\"\i ve form,
one should first `dress' the theory
by tensoring it with an extra $U(1)$
current. The extra degrees of freedom
which thus appear are then killed by
projecting onto a special form of
the null state. Therefore, although the
null vector in the matter sector is the same, the specific form of the
equation for correlation functions changes.

We begin in the next section by introducing the KP \h\ and the
notion of Kon\-tse\-vich-Miwa transform. Then, in Sect.~3, we
 introduce the dressed \de. For purely technical reasons, we describe
 in this paper the coincidence between W constraints and \de s
 proceeding in the direction
{\it from\/} the \de s {\it to\/} the
W \cs: we show in Sect.~4 how to
 `unkontsevich' conformal field-theoretic data into the \w3
 constraints. It will be clear from the derivation that (and how) one
can reverse the argument and
thus establish {\it equivalence\/} of the
 two sets of data, the \w3-constrained KP \h\ and ({\it appropriately
 dressed\/}) minimal models.

\section{The
KP \h\ and the \K-\M\ transform}\lvm We start with the KP \h\
\cite{[DDKM]}. It can be described either
in an evolutionary form, as an
infinite set of equations on
(coefficients of) a pseudodifferential operator,
or as (Hirota) bilinear relations
on the tau function. The tau-functional
description may be considered as having the more direct relevance to
`physics', being related to the
partition function, while the evolutionary
form has all the usual advantages
due to the introduction of a spectral
parameter and the associated wave function.
The wave function and the adjoint
wave function depend on the spectral parameter $z$ via
$\exp(\pm\xi(t,z))\tau(t\mp[z^{-1}])/\tau(t)$, where
$\xi(t,z)=\sumr t_rz^r$
and \BE t\pm [z^{-1}]=\left(t_1\pm z^{-1},
t_2\pm\half z^{-2}, t_3\pm {1\over 3}
z^{-3},\ldots\right)\ .\label{timesshift}\EE Here
$t\!=\!(x\!\equiv\!t_1,t_2,t_3,\ldots)$ are the time
parameters of the \h.
This form of the $z$ dependence is `generalized' by the following
presentation for the $t_r$, known as the \M\ parametrization
\cite{[Mi],[Sa]}: \BE t_r={1\over r}\sum_j n_jz^{-r}_j,\quad
r\geq1\label{Miwatransform}\EE where $\{z_j\}$ is a
set of points on the
complex plane and the \M\ parameters
$n_j$ are integer classically; we will
need, however, to continue off the integer
values. As we will see, the Miwa
parameters acquire the r\^ole of
$U(1)$ charges w.r.t. the scalar field.

By the \K-\M\ transform we will understand the dependence, via
eq.\req{Miwatransform}, of $t_r$ on the
$z_j$. However, it is important that
the presentation of the form
\req{Miwatransform} be considered for arbitrary
$n_j$ (while, on the other hand,
\K\ has originally used a parametrization of
this type for all the $n_j$ equal \cite{[K]}, see also [15-18].

As was noted in the Introduction, an important result of the
development of \mm s was the
discovery of an `integrable' counterpart of
the 2$d$ gravity~+~matter theories, in
the form of \V-{\it constrained\/}
\hs. Moreover, the fact that appropriately constrained \hs\ provide a
description of 2$d$ gravity coupled to matter, may be promoted to a
first-principle, in which case it is
natural to consider on an equal footing
various \cs\ more general than the \V\ ones, including those whose
derivation from a specific {\it matrix\/}
model is not known. This applies,
first of all, to the extension of \V\ \cs\ to W-ones, and also to the
 introduction of a parameter
 into the \V\ generators acting on the tau function \cite{[S2912]}.
 This parameter $J$ (which would have
parametrized the central charge as $2(6J^2-6J+1)$, had the
 ${\Sc L}_{n\leq-2}$ generators been
present as well),
can be thought of as the `spin' (dimension) of an abstract
$bc$ system underlying the \V\ generators, $\sum_{n\in{\open Z}}{\Sc
L}_nz^{-n-2}\sim (1-J)\d bc-Jb\d c$.
We repeat that although $J$-dependent \V\ \cs\ may not have been
derived from a {\it matrix\/} model, for us all of them are
equally good starting points.
 The  corresponding \cs,
\BE{\Sc L}_n\tau=0,\quad n\geq-1,\EE
\BE\new \BA{rcl} {\Sc L}_{p>0} &=&\half\sum^{p-1}_{s=1}{\d^{2}\over\d
t_{p-s}\d t_s}+\sum_{s\geq 1}st_s {\d\over\d t_{p+s}}+
\left(J-\half\right)(p+1)
{\d\over\d t_p}\\ {\Sc L}_0&=&\sum_{s\geq
1}st_s {\d\over\d t_s}\\ {\Sc
L}_{-1}&=&\sum_{s\geq 1}(s+1)t_{s+1}{\d\over\d t_s} \  ,
\EA\label{Lontau}\EE
 were shown in \cite{[S35]} to take, after the \K-\M\ transform,
the form of a decoupling equation
corresponding to the level-2 null vector in
 the $(p\pr,p)$ minimal model determined by
 $p\pr/p=2n_i^2$, provided the \M\ parameter $n_i$
was determined from\BE{1\over n_i}-2n_i=2J-1\equiv Q\qquad\qquad{\rm
(\V)}\ . \label{ni}\EE

Our task in this paper is to extend this correspondence to include W
generators. Let us start therefore with
the \w3 \cs\ imposed on the tau
function in a way similar to \req{Lontau}; we will write out the \cs\
explicitly later, see \req{w3generators}. Now, by virtue of the \K-\M\
transform \req{Miwatransform}, the tau function becomes a function
$\tau\{z_j\}$ of the $z_j$. We
assume for it the ansatz\footnote{Clearly,
 there have to be infinitely many points
 $z_j$ in the \K-\M\ transform in order
 for the times $t_r$ to be independent.}
\BE\tau\{z_j\}=\lim_{n\rightarrow\infty}\langle\Psi
(z_1)\ldots\Psi(z_n)\rangle\label{ansatz}\EE with
$\langle~~\rangle$ and
$\Psi$ being, respectively, the chiral
correlation function and a (primary)
field operator in a \cft\ on the
$z$ plane. The operator and the theory
itself should be specified according to the fact that the tau function
satisfies the \w3 \cs.
 The ansatz will be justified by the fact that,
 as we are going to show, imposing these \cs\ amounts to
 the condition that $\Psi$ be a `31'
operator in a $(p\pr, p)$ minimal model \cite{[BPZ],[DF]} for $p\pr/p$
determined by the precise form of the
\cs. That is, we will find the relation
$p\pr/p=\half n_i^2$, where $n_i$ is the
Miwa parameter which in its own turn
is related to a `spin' $J$, read
off from the {\it Virasoro part\/} of the
\cs, by\BE{2\over
n_i}-n_i=2J-1\qquad\qquad {\rm(W}^{(3)}{\rm)}\label{niW3}\ . \EE

As was noted above, we find
it more convenient to present in this paper
 the derivation of the correspondence W
 constraints~$\leftrightarrow$~decoupling equations in the inverse
direction: we start from $\Psi$ being the `31' operator and then show
how to `unkontsevich' it into the \w3 constraints. We therefore
proceed in the next section with introducing the necessary conformal
field-theoretic ingredients.

\section{`Dressing' the decoupling equation}\lvm Now we introduce
the second ingredient: null-vector \de s in minimal models of
conformal field theory. To the KP \h\ we will return at the very end.

In addition to the \emt\
$T(z)=\sum_{n\in {\open Z}}L_nz^{-n-2}$, consider a
$U(1)$ current $j(z)=\sum_{n\in{\open
Z}}j_n z^{-n-1}$. The commutation
relations read, in the
standard manner,\BE\new\BA{rcl} \L[ j_m,j_n\R] &=&
m\delta_{m+n,0}\\ \L[L_m,L_n\R]&=&(m-n)L_{m+n}+{d+1\over 12}(m^3
-m)\delta_{m+n,0}\\
\L[L_m,j_n\R]&=&-nj_{m+n}\ . \EA\label{thetheory}\EE The
central charge is parametrized as $d+1$, with 1 being the $U(1)$
contribution.

Now, let $\Psi$ be a primary
field with conformal dimension $\Delta$ and
$U(1)$ charge $q$. We are interested
in null vectors at level 3 \cite{[BPZ]}.
It is straightforward to see that a
null vector will result from the action
on $|\Psi\rangle$ of the following operator:\BE\new\BA{c} \delta
L_{-3}-2L_{-2}L_{-1}+{1\over\delta+1}L_{-1}^3
+{\delta+1+3q^2\over\delta+1}j_{-1}^2L_{-1}
+q{2\delta-1\over\delta+1}j_{-2}L_{-1}+2qj_{-1}L_{-2}\\
- {3q\over\delta+1}j_{-1}L_{-1}^2 -{\delta^2+\delta+2\delta
q^2-q^2\over\delta+1}j_{-1}j_{-2}
-q{\delta+1+q^2\over\delta+1}j_{-1}^3
-q{\delta(\delta-1)\over\delta+1}j_{-3}\EA\label{decouplingopfull}\EE
where
\BE\delta=\Delta-{q^2\over 2},\EE provided
\BE\delta={7-d\mp\sqrt{(1-d)(25-d)}\over 6}\label{delta}\ . \EE
(It is understood
that for the $(p\pr,p)$
minimal model, $d=1-{6(p\pr-p)^2\over p\pr p}$.)

The operator \req{decouplingopfull} allows too much arbitrariness,
as the values of $\Delta$ (or
$\delta$, which is the {\it matter\/} dimension
of $\Psi$) and $q^2$ cannot be fixed separately. However, the \K-\M\
 transform suggests
a more special form of the operator \req{decouplingopfull},
 which can be arrived at as follows.
One first converts the expression
 of the null vector into the \de\ for correlation
functions of the form\BE\L\langle\Psi(z_i)\prod_{j\neq i}{\cal
O}_j(z_j)\R\rangle\ . \EE where, obviously, at
least one operator insertion must
be that of $\Psi$. (From now
on, the corresponding insertion point $z_i$
will therefore be singled out from the rest of the $z_j$. The other
insertions, of dimensions\footnote{Dimensions refer
to those as evaluated in
the combined theory, by operator
product expansions with the \emt\ $T(z)$. On the other hand,
separating away the
current contribution by writing $L_n=l_n+({\rm
Sugawara})_n$, would leave us with the
{\it matter\/} sector in which the
dimensions are found from the \emt\
composed of the $l$'s.} $\Delta_j$ and
$U(1)$ charges $q_j$, may or may not coincide with $\Psi$).

Then, analysing the \de\ corresponding to \req{decouplingopfull},
one finds that
the terms coming from $j_{-1}^3$ cannot be recovered from the \K-\M\
transform of W \cs\ on the KP \h\footnote{
The terms $$\sumji\sum_{k\neq i}\sum_{l\neq
i}{q_jq_kq_l\over(z_j-z_i)(z_k-z_i)(z_l-z_i)}
\L\langle\Psi(z_i)\prod_{j\neq
i}{\cal O}_j(z_j)\R\rangle,$$ corresponding to
$j_{-1}^3$, are {\it not\/} in
the image of the \K-\M\ transform of the W constraints.}.
 To kill these terms, we have to choose
\BE\delta=-1-q^2\label{relation}\ . \EE

Further, with this condition satisfied, there
are terms in the \de\ of the
form\BE\sum_{{j,~\!k\atop j\neq i,~\!k\neq i}}\!{q_j\over
(z_i-z_j)(z_k-z_i)^2} \L\{-(q^2+2)q_k-2q\Delta_k\R\}\L
\langle\Psi(z_i)\prod_{j\neq i}{\cal O}_j(z_j)\R\rangle\ , \EE
which are not
acceptable either. Thus, similarly to
the \V\ case \cite{[S35]}, we restrict
ourselves to a subsector of
those operators whose dimensions and $U(1)$
charges
satisfy\BE\Delta_j=-\half(q^2+2){q_j\over q}\ . \label{dressing}\EE
This implies fixing a {\it dressing\/} prescription: the extra scalar
$\phi(z)\sim\int^z\!j$ enters vertex operators
with an exponent depending on
that of the matter part.

As a result of conditions \req{relation} and \req{dressing}, the \de\
takes a very
simple form\BE\new\BA{rcl}\Biggl\{-{1\over q^2}{\d^3\over\d
z_i^3} &+&\sumji{1+q^2\over(z_j-z_i)^2} \L(\dd{z_j}-{q_j\over
q}\dd{z_i}\R)+\sumji{1\over z_j-z_i}\L(2
{\d^2\over\d z_j\d z_i}-3{q_j\over
q}\ddsc{z_i}\R)\\
{}&+&2\sumji\sum_{k\neq i}{q_k\over(z_j-z_i)(z_k-z_i)}
\L(q\dd{z_j}-q_j\dd{z_i}\R)\Biggr\}\L\langle\Psi(z_i)\prod_{j\neq
i} {\cal
O}_j(z_j)\R\rangle=0\ .\EA\label{decoupling}\EE

In the chosen normalization (+ sign on
the RHS of the first equation in
\req{thetheory}), the $q$'s must be
imaginary ($q^2<0$), so let us define
\BE\L\{\BA{rcl} q_j&=&\sqrt{-1}n_j,\quad
j\neq i\\ q&=&\sqrt{-1}n_i\EA\R.
\label{q}\EE We claim that these are {\it the\/} $n$'s from the \K-\M\
transform! In the next section we show how to `unkontsevich'
eq. \req{decoupling} into the full set of \w3 \cs.

Note that in a similar analysis
of the level-2/\V\ case \cite{[S35]}, one
obtains at this stage a generalized {\it master equation\/} (cf.
\cite{[MS]})\BE\L\{-{1\over 2n_i^2}{\d^2\over\d z_i^2}+{1\over
n_i}\sum_{j\neq i}{1\over z_j-z_i}\L(n_j{\d\over\d z_i}-n_i{\d\over\d
z_j}\R)\R\} \L\langle\widetilde{\Psi}(z_i)\prod_{j\neq i} {\cal
O}_j(z_j)\R\rangle=0\label{Tgen}\EE in which
$\widetilde{\Psi}$ is the `21'
operator. The null vectors in the
{\it matter\/} sector, underlying each one
of these equations, are of course
the same as for the respective standard
\de s \cite{[BPZ]}. However, the specific
form that the \de\ takes in the presence
of the current, is crucial for the derivation of W generators.

\section{From the decoupling
 equation to the \w3 \cs}\lvm Having arrived in the
 last section at the {\it decoupling operator}
\BE\new\BA{rcl}-n_i{\cal W}\equiv&{}&{}\\{1\over
n_i^2}{\d^3\over\d z_i^3}
&+&\sumji{1-n_i^2\over (z_j-z_i)^2} \L(\dd{z_j}-{n_j\over
n_i}\dd{z_i}\R)+\sumji{1\over z_j-z_i}\L(2
{\d^2\over\d z_j\d z_i}-3{n_j\over
n_i}\ddsc{z_i}\R)\\
{}&-&2\sumji\sum_{k\neq i}{n_k\over(z_j-z_i)(z_k-z_i)}
\L(n_i\dd{z_j}-n_j\dd{z_i}\R)\ . \EA\label{decouplingop}\EE let
us now interpret
the $z_j$ and $n_j$ as the ingredients of the \K-\M\ transform
\req{Miwatransform}, and see whether the
operator ${\cal W}$ can indeed be
rewritten in terms of the time parameters $t_r$. This is by no means
automatic, and in particular would not be true had we not imposed the
restrictions \req{relation} and \req{dressing}!

A trivial part will be to express
${\d/\d z}$'s in terms of ${\d/\d t}$'s.
The non-trivial part is to get rid of the $z_j$ themselves, as
eq. \req{Miwatransform} does not allow
this to be done straightforwardly.
As an example let us take
the second term from \req{decouplingop}. Using
\req{Miwatransform} to express derivatives w.r.t. the $z_j$, we find
\BE\new\BA{l} \sumji{1-n_i^2\over (z_j-z_i)^2} \L(\dd{z_j}-{n_j\over
n_i}\dd{z_i}\R)= \sumji{1-n_i^2\over z_j-z_i}n_j\sumr
{z_i^{-r-1}-z_j^{-r-1}\over z_j-z_i}\dd{t_r}\\
=-(1-n_i^2)\sum_{p\geq-1}\!\!\sum_{{q\geq1\atop q+p\geq1}}
\!\!(p+2)z_i^{-p-3}qt_q\dd{t_{p+q}}\\
+(1-n_i^2)n_i\sum_{p\geq1}\half(p+1)(p+2)z_i^{-p-3}\dd{t_p}
+(1-n_i^2)\sumji{n_j\over
z_j-z_i}\!\sum_{p\geq1}z_i^{-p-2}(p+1)\dd{t_p}
\ . \EA\EE

Similarly, in the last term in
\req{decouplingop} we can also divide by
$z_j-z_i$ in the following
way: \BE\new\BA{l} -2\sum_{k\neq i}{n_k\over
z_k-z_i}\sumji n_in_j\sumr
{z_i^{-r-1}-z_j^{-r-1}\over z_j-z_i}\dd{t_r}\\
=-2n_i\sumji{n_j\over
z_j-z_i}\sum_{p\geq-1}\!\sum_{{s\geq1\atop p+s\geq1}}
 st_sz_i^{-p-2}\dd{t_{p+s}}+2n_i^2\sumji{n_j\over
 z_j-z_i}\sum_{p\geq1}(p+1)z_i^{-p-2}\dd{t_p}\EA\EE

 Computing in a similar manner
 the terms with the second and the third derivatives and collecting
everything together, we get\BE\new\BA{rcl}{\cal
W}&=&\sum_{q,r,s\geq1}z_i^{-q-r-s-3}{\d^3\over\d t_q\d
t_r\d t_s}+\L({3\over
n_i}-2n_i\R)\sum_{r,s\geq1}z_i^{-r-s-3}(r+1){\d^2\over\d t_r\d t_s}\\
{}&+&\L({1\over
n_i^2}+{n_i^2\over2}-\half\R)
\sumr(r+1)(r+2)z_i^{-r-3}\dd{t_r}+{1-n_i^2\over
n_i}\sum_{p\geq-1}\!\!\!\sum_{{q\geq1\atop q\geq1-p}}\!z_i^{-p-3}
(p+2)qt_q\dd{t_{p+q}}\\
{}&+&2\sum_{p\geq0}z_i^{-p-3}\sum_{s=-1}^{p-1}\!\!\sum_{{q\geq1\atop
q\geq1-s}}\!qt_q {\d^2\over\d t_{p-s}\d
t_{q+s}}+2\sumji {n_j\over z_j-z_i}
\sum_{p\geq-1}{\Sc L}_p\L(z_i^{-p-2}-z_j^{-p-2}\R)\\
{}&{}&\hfill+2\sumji{n_j\over z_j-z_i}
\sum_{p\geq-1}{\Sc L}_pz_j^{-p-2}
\EA\label{LL}\EE with \BE{\Sc
L}_{p\geq1}=\half\sum^{p-1}_{s=1}{\d^2\over\d t_{p-s}\d
t_s}+\sum_{s\geq1}st_s\dd{t_{p+s}}+
{2-n_i^2\over2n_i}(p+1)\dd{t_p}\label{Lbar}\EE and   ${\Sc L}_0$ and
${\Sc L}_{-1}$ are the same as
in \req{Lontau}. Notice that the \V\ \cs\
(4.5) are different from the ones in (2.4), according to the value
of $J$ in terms of the Miwa parameter (assuming it is the same).
The last term has
been added to and subtracted from
the RHS of \req{LL}. With the combination
$(z_i^{-p-2}-z_j^{-p-2})/(z_j-z_i)$ we proceed as above,
and in this way we
finally arrive at,\BE 0={\cal W}\tau=\sum_{p\geq-2}z_i^{-p-3}{\Sc
W}_p^{(3)}\tau+2\sumji{n_j\over z_j-z_i}\sum_{p\geq-1}z_j^{-p-2}{\Sc
L}_p\tau\label{linearcomb}\EE with
\BE\new\BA{lclclcl}{\Sc
W}_{-2}^{(3)}&=&2\sumr rt_r{\Sc L}_{r-2}&{}&{}&{}&{}\\ {\Sc
W}_{-1}^{(3)}&=&2\sumr rt_r{\Sc L}_{r-1}&-&2n_i{\Sc
L}_{-1}&+&{1-n_i^2\over n_i}\sum_{r\geq2}rt_r\dd{t_{r-1}}\\
{\Sc W}_0^{(3)}&=&2\sumr rt_r{\Sc L}_r&-&4n_i{\Sc
 L}_0&+&2{1-n_i^2\over n_i}\sumr
rt_r\dd{t_r}~~+~~2\sum_{r\geq2}rt_r{\d^2\over\d t_1\d t_{r-1}}\\ {\Sc
W}_{p\geq1}^{(3)}&=&2\sumr rt_r{\Sc L}_{r+p}&-&2(p+2)n_i {\Sc
L}_p&+&(p+2){1-n_i^2\over n_i}\sumr rt_r\dd{t_{r+p}}\\
{}&{}&{}&{}&{}&{}&\hfill+~~2\sum_{s=-1}^{p-1}\!\!\sum_{{r\geq1\atop
r\geq1-s}}rt_r{\d^2\over\d t_{p-s}\d t_{r+s}}\\
{}&+&\multicolumn{3}{c}{(p\!+\!1)(p\!+\!2)\!\L(\!{1\over
n_i^2}\!+\!{n_i^2\over2}\!-\!\half\!\R)\!\dd{t_p}}&+&\L(\!{3\over
n_i}\!-\!2n_i\!\R)\!\sum_{r=1}^{p-1}(r\!+\!1){\d^2\over\d
t_r\d t_{p-r}}
+\!\!\!\!\sum_{{q,r,s\geq1\atop q+r+s=p}}\!\!\!\!{\d^3\over\d
t_t\d t_r\d
t_s}\EA\label{w3generators}\EE

The final step consists in demanding that the ${\Sc W}^{(3)}$ and the
 ${\Sc L}$ generators annihilate the tau function
 separately\footnote{according to the form of the dependence
on $z_i$ and $z_j$ in
\req{linearcomb}; in particular, the combination
$\sumji{n_j\over z_j-z_i}$ can{\it not\/} be
rewritten in terms of the time
parameters, while all the rest
of eq.\req{linearcomb} {\it is\/} expressed
through the times, so that the
two terms (those involving ${\Sc W}^{(3)}$ and
${\Sc L}$ respectively) are `linear
independent', hence the vanishing of
 each one.}:\BE{\Sc L}_n\tau=0,\quad n\geq-1,\qquad{\Sc
W}_n^{(3)}\tau=0,\quad n\geq-2.\EE Thus we have
recovered the full set of \w3
\cs, (i.e. ${\Sc W}^{(3)}_n$ and
${\Sc L}_n$) with the coefficients in
 front of the different terms depending on $n_i$ and therefore, via
eq.\req{niW3}, on the corresponding minimal model. Recall also that,
according to \req{relation}, \req{dressing},
\req{delta} and \req{q}, the
Miwa parameter $n_i$ is equal to the cosmological constant\BE
n_i=\sigma\a_{\pm}\equiv\sigma\L(-\half\sqrt{{25-d\over3}}\pm\half
\sqrt{1-d\over3}\R),\EE where $\sigma^2=1$ and the upper/lower sign
corresponds to that in \req{delta}.
Stated differently, for the $(p\pr,p)$
minimal model we have $n_i^2=2p\pr/p$.

Conversely, {\it starting\/} from the full
set of \w3 \cs, one constructs the
combination \req{linearcomb} (however unnatural it
may seem from the point of
view of the $t$-variables) and then,
fixing the coefficients as explained and
inverting the previous steps, one arrives at the \de. This proves the
equivalence between the \w3 \cs\ and
the level-3 \de\ in the $(p\pr,p)$
 minimal model.

A remark is in order concerning why
the \V\ part of the derived constraints
does not lead to the level-2 \de
s as was the case in ref. \cite{[S35]}.
The point is that the possibility to
transform the \V\ \cs\ into a \de\
depends crucially on the relation between
the `spin' $J$ present in the \V\
generators, and the value of the Miwa parameter $n_i$ (recall that the
$i$-{\it th\/} point, being the position of the insertion of
$\Psi=\Psi_{31}$, plays a special r\^ole). However,
the spin $J$ as read off
from the generators \req{Lbar},
eq. \req{niW3}\footnote{Note that it follows
from \req{niW3}
that $J-\half=(\pm)\half\sqrt{(1-d)/3}\equiv\half(\pm)Q_{{\rm
m}}$, where $Q_{{\rm m}}$ is
the matter background charge, which therefore
coincides with $Q\equiv2J-1$.}, is {\it not\/} the same as required by
\req{ni} and therefore the generators \req{Lbar} do not allow a
transformation into the $z$-variables.

\section{Dressing prescription and generalizations}\lvm

The dressing prescription \req{dressing} corresponding
to \w3 \cs, can be
rewritten as $\Delta_j=\pm\half\sigma Qn_j$ and
in this form it coincides
with the dressing prescription derived from the \V\ \cs\ on the KP \h\
\cite{[S35]}. Therefore the sector singled
out from the matter$\otimes U(1)$
theory by the dressing prescription is
the same for the two cases. Using
another `universal' relation,
$\delta_j=\Delta_j+\half n_j^2$, which gives
the matter dimensions by subtracting away
the Sugawara part, we arrive at\BE
n_j^2\pm\sigma Qn_j-2\delta_j=0.\EE The value of $n_j$ found from here
differs from the DDK dressing by the cosmological constant $\a_{\pm}
=\half(-Q_{{\rm L}}\pm Q)$.

A general pattern which generalizes the two
lowest cases, \w2 and \w3, can be
formulated as follows: To the \w{l} \cs\
on the KP \h\ we associate the \de\
corresponding to the irreducible module built on the ({\it dressed\/})
$(l,1)$ (or $(1,l)$) state in the
$(p\pr,p)$ minimal model. The value of
$p\pr/p$ is optional, and depends on the
`spin' $J$ entering in the {\it \V\
part\/} of the \w{l} \cs:\BE{p\pr\over
p}=1+{Q^2\over4}+{Q\over4}\sqrt{Q^2+8},\qquad Q\equiv2J-1.\EE
According to
the standard formulae, the primary
field $\Psi_{l1}$ has dimension (in the
matter
sector) \BE\delta={4+Q^2+Q\sqrt{Q^2+8}\over16}(l^2-1)+{1-l\over2}\ . \EE
The dressing of $\Psi_{l1}$ is specified by its $U(1)$ charge
$q=\sqrt{-1}n_i$ with $n_i$, which at the
same time is the Miwa parameter,
determined from \BE\delta={l+1\over l-1}{n_i^2\over2}+{1-l\over2}\EE
(eq. \req{relation} is recovered for $l=3$). This gives for the `bulk'
dimension of the dressed $\Psi_{l1}$
operator:\BE\Delta=\delta-{n_i^2\over2}={n_i^2\over
l-1}+{1-l\over2}.\EE The
dimensions and $U(1)$ charges of all the other operators
satisfy\BE\Delta_j=\Delta{n_j\over n_i}=\L({n_i\over
l-1}-{l-1\over2n_i}\R)n_j\label{z}\ . \EE With $n_i$
determined from the above
formulae as\BE n_i=(1-l){Q+\sqrt{Q^2+8}\over4},\EE
we find the relation
between the Miwa parameter $n_i$ and the
spin $J$ present in the \V\ part of
the \w{l} \cs:\BE{l-1\over n_i}-{2n_i\over l-1}=2J-1\qquad\qquad
{\rm(W}^{(l)}{\rm)}\ . \EE For the $(1,l)$
operator the formulae would receive
the obvious sign modifications.

\bigskip

To conclude let us note that beyond level
4 it would be difficult to give a
direct derivation of the correspondence
between \w{l} \cs\ and the $(l,1)$
\de s, so one has to rely on more general arguments (cf. however,
ref. \cite{[BSa]}). Also, it would be
interesting to understand the r\^ole in
this picture of the {\it W-algebra\/} null
vectors, as well as of the \de s
other than those considered in \cite{[BSa]}.
It is also unclear how {\it
integer\/} values of $p\pr$ and $p$ could
be singled out in the KP formalism.
These problems are under investigation and
will be the subject of a future
publication.

\bigskip
{\bf {\sc Acknowledgements}}. We would like to thank
A.~Berkovich, V.~Fock, E.~Kiritsis and W.~Lerche
for useful discussions.

\end{document}